# Modélisation de la propagation dans une structure fermée surdimensionnée

F. DEMONTOUX, F. BONNAUDIN et J.L. MIANE

Laboratoire POIM (Physique des Interactions Ondes-Matière)
ENSCPB  16 avenue Pey-Berland  33607 PESSAC cedex

## I- Introduction

La propagation d'une onde EM dans des structures fermées conduit à de nombreuses applications allant de la caractérisation de matériaux dans des cavités résonantes aux mesures en enceintes réverbérantes utilisées en CEM. Nous avons étudié à partir de modélisations, les différentes conditions d'existence des ondes électromagnétiques dans des structures fermées surdimensionnées. Deux modes peuvent exister dans de telles structures. Un de ces modes repose sur un fonctionnement en continu avec des modes résonnants particuliers dont on peut tirer parti pour diverses applications dont nous donnerons des exemples. Le deuxième mode consiste en un fonctionnement pulsé qui s'apparente à la propagation des ondes radars et qui apporteront d'autres informations.

Les structures que nous présenterons sont fermées par des parois parfaitement conductrices et l'onde électromagnétique est générée dans la structure par des antennes dipôles. Les dimensions (longueur, largeur et hauteur) sont toujours plus grandes que la longueur d'onde, le facteur de surdimensionnement restant de l'ordre de dix.

## II- Fonctionnement en CW : enceintes multimodes

L'enceinte est équipée de deux embases coaxiales qui permettront d'introduire l'onde dans la structure. Ces antennes sont modélisées ainsi que les parois conductrices. Les calculs des répartitions de champ sont effectués à l'aide du logiciel HFSS de la société ANSOFT qui utilise la méthode des éléments finis afin de résoudre les équations de Maxwell. La structure ainsi créée se comporte comme une cavité résonnante. Une première modélisation permet de déterminer les fréquences de résonance correspondant à des modes particuliers. Les figures suivantes donnent un exemple d'une structure et d'un mode résonnant particulier.

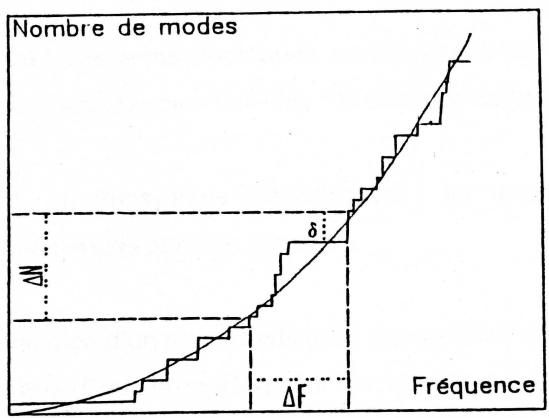

*Figure 1 : Nombre de modes de fréquence de résonance inférieure à F.*

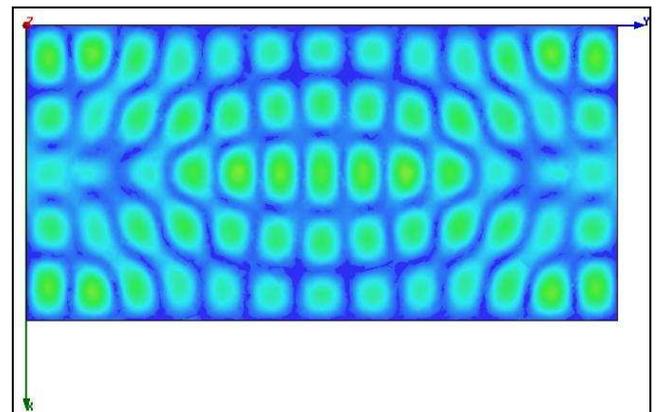

*Figure 2 : Exemple d'un mode résonnant à 2,450 GHz
(1000 mm * 500 mm * 100 mm)*

Les différents modes résonnants sont proches, ils peuvent être dénombrés pour des enceintes de forme simples, et il apparaît que sur un intervalle de fréquence ΔF un grand nombre de modes peuvent être excités. Si on génère dans l'enceinte des ondes électromagnétiques sur cet intervalle de fréquence, soit simultanément soit successivement par un balayage en fréquence, nous obtenons un effet de superposition. Le champ moyen est pratiquement uniforme dans l'enceinte. C'est ce que tente de réaliser les « brasseurs d'ondes » des fours micro-ondes ménagers. Il est aussi possible d'utiliser cette particularité pour des applications en caractérisation [1].

### III- Fonctionnement pulsé

Un code de calcul basé sur la méthode FDTD [1] développé au laboratoire PIOM nous a permis d'étudier ce type de fonctionnement. Un pulse très court de quelques nanosecondes a était généré dans notre structure par simulation au niveau de l'antenne. La structure modélisée est la même que la précédente, l'antenne émettrice est placée à 92 mm sur l'axe longitudinal de l'enceinte. Une second antenne est placée à l'autre extrémité de l'enceinte toujours à 92 mm du bord. Les signaux obtenus en réflexion et en transmission sont calculés sur un intervalle de temps de l'ordre de la dizaine de nanosecondes. Au-delà nous retrouvons l'établissement d'un mode résonnant (si la fréquence convient) ou l'absorption de l'impulsion pour les fréquences qui ne peuvent pas résonner. En comparant les signaux reçus lorsque l'enceinte est vide et lorsqu'elle contient un échantillon de matériau, il apparaît que ces signaux peuvent permettre de localiser l'échantillon (mesure en réflexion, comme le ferait un radar) et de mesurer son absorption (mesure en transmission). Les figures suivantes 3 et 4 donnent un exemple de l'amplitude du champ détecté, en réflexion sur la première antenne, en transmission sur la seconde antenne.

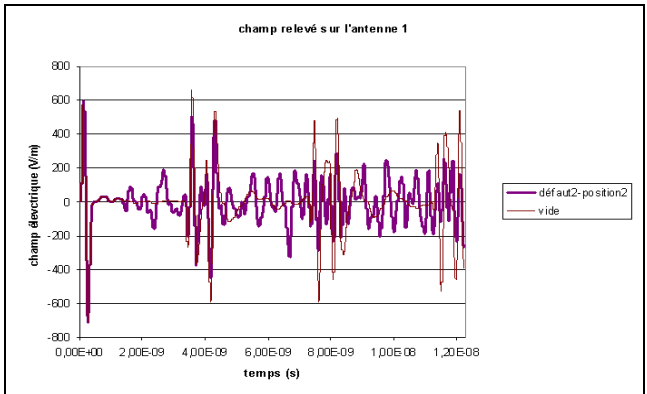
*Figure 4 : Signal temporel obtenu en réflexion*

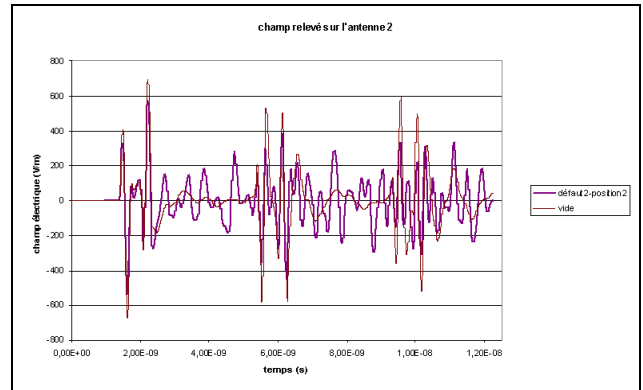
*Figure 5 : Signal temporel obtenu en transmission*

L'analyse des données dans le domaine temporelle (détection du défaut, recherche de sa position) nécessite une bonne connaissance des phénomènes qui se déroulent à l'intérieur de la cavité. Pour cela nous avons décomposé notre source électromagnétique isotrope (dans un plan) en deux sources anisotropes. La superposition des deux sources ainsi créées représente la source d'origine (figures 6, 7 et 8). Chacune des deux sources possède une direction de propagation privilégiée suivant l'axe longitudinal de la cavité mais dans des directions opposées. L'étude des données dans le domaine temporelle peut alors être décomposé selon les sources utilisées (figures 9,10 et 11). Il est alors possible d'identifier l'origine des signaux temporels observés (réflexion sur une paroi, réflexion sur un échantillon…).

### IV-Conclusion

Nous avons modélisé la propagation d'onde électromagnétique dans une enceinte fermée aux temps courts, avant l'établissement des modes résonnants, et au temps longs où les résonances sont observées. La connaissance du comportement EM des enceintes fermées permet d'envisager des techniques de contrôles non destructifs de l'état de ces enceintes et des techniques de recherche de défaut dans des structures fermées.

### Références

[1] Etude et réalisation d'un dispositif d'analyse diélectrique et thermique utilisant un résonateur multimode à balayage
S. YACOUBI Thèse de l'Université de Bordeaux 1989